\documentstyle[12pt]{article}
\begin{document}
\thispagestyle{empty}
\def\baselinestrecht{1.3}
\rightline{\bf UG-FT-22/92}
\rightline{\bf May 1992 \ \ }
\vskip 0.6cm
\begin{center}
{\bf
$Z'$ Decays into Four Fermions $^*$
\\}
\vskip 1.5cm
\small
{\bf F. del Aguila, B. Alles} \\
{\it
Departamento de F\'{\i}sica Te\'orica y del Cosmos, \break
Universidad de Granada, E-18071 Granada, Spain \\}
\vskip 0.5cm
{\bf Ll. Ametller} \\
{\it
Departament de F\'{\i}sica i Enginyeria Nuclear,\break
Universitat Polit\`ecnica de Catalunya, E-08028 Barcelona, Spain \\}
\vskip 0.5cm
and \\
\vskip 0.2cm
{\bf A. Grau $^\dagger$}\\
{\it
INFN, Laboratori Nazionali di Frascati, \break
P.O. Box 13, 00044 Frascati, Italy \\}
\vskip 2.cm
\end{center}
\begin{abstract}
 If a new $Z'$ is discovered with a mass $\sim 1 \ TeV$ at LHC/SSC,
its (rare) decays into two charged leptons plus missing transverse energy
will probe the $Z'$ coupling to the lepton doublet
$\left(
\begin{array}{c}
\nu \\
e
\end{array}
\right)_L$ and to $W^+W^-$,
allowing further discrimination
among extended electroweak models.
\end{abstract}
\vskip 1cm
\noindent
--------------------------------------- \break
{\small
{$^*$ Work partially supported by CICYT under contract
AEN90-0683} \hfil \break
{$^\dagger$ On leave of absence from
Departamento de F\'{\i}sica Te\'orica y del Cosmos,
Universidad de Granada, E-18071 Granada, Spain}
}

\newpage
\setcounter{page}{1}
\normalsize

 Many approaches to extending the standard model lead to a larger
gauge group which implies a new heavier $Z'$ boson which some day
will be discovered. Then it will be important to learn experimental
techniques to use the $Z'$ decays to determine what the new gauge
group is. In this letter we extend the analysis of this question.

 Large hadron colliders offer the best chance to observe (in contrast with
indirect evidence) a new heavy $\sim 1 \ TeV$ gauge boson, $Z'$
\cite{ZP,REVIEWZP,LHC}.
If it couples to quarks and leptons with a sizeable strength ($g\sim 0.1$)
it should be discovered in the two lepton final states, $e^+e^-$ and
$\mu^+\mu^-$. The subsequent measurement of the forward-backward
asymmetry should constrain the ratio of the axial to the vector charged
lepton $(Z')$ couplings,
discriminating among possible extended electroweak models.
Recently it has been pointed out that the $Z'$ decays into two fermions
and a $W$ or a $Z$ would further distinguish among models \cite{LANGACKER}.
These processes, however, appear in detectors
as four fermion final states and
then require a definite strategy for their identification.
As a matter of fact, the $Z'$
decays into $W^+W^-$ also contribute to this signal.
We argue
that for the popular ($E_6$) models the most interesting
decays are those
involving two charged leptons and two neutrinos, although the cross section
is small \cite{WW}. These decays constrain the
$Z'$ coupling to the lepton doublets
$\left(
\begin{array}{c}
\nu \\
e
\end{array}
\right)_L$
and
$\left(
\begin{array}{c}
\nu \\
\mu
\end{array}
\right)_L$
as well as the $Z'W^+W^-$ coupling,
and then do discriminate among models.
Two samples can be distinguished, one containing the
events with two charged leptons of different flavor, $e^-\mu^+\rlap/p$
and $e^+\mu^-\rlap/p$,
and other the events
with two charged leptons of the same flavor,  $e^-e^+\rlap/p$
and $\mu^-\mu^+\rlap/p$. These are final products of $Z'$ decays
into $f\bar f W$,
$f\bar f Z$ and $WW$, where $f$ can be $e, \mu $ or $\nu$.
Both samples are of the same size and give similar information
but the first one is cleaner (and simpler to evaluate).
Elsewhere we discuss
the sample with two charged leptons of the same flavor and
the other four fermion channels. The latter, however, have too small
a cross section (four charged leptons)
or too large a background (non leptonic final states).

 Let us present first the numerical
results (plots) and postpone the discussion to the
end. We assume that a new $Z'$ with a mass, for instance,
of $1 \ TeV$ is known to exist.
And we are interested in events with one electron and one muon
($e^-\mu^+$ or $e^+\mu^-$) plus missing
transverse energy ($\rlap/p _t$).
The two
main backgrounds result from the $WW$ continuum and from
heavy quark ($t$) production. Whereas the first is irreducible (and
after cuts small), the second is difficult to estimate. We assume that the
latter will be controlled once large transverse momenta for
$\rlap/p$ and/or $e, \mu$ are required and
a good understanding of the heavy
quark ($t$) production is obtained,
using criteria of isolation and multiplicity
\cite{LHC}.
We concentrate on the four fermion $Z'$ signal and the
standard model (continuum) $WW$ background.
The former gets contributions from five diagrams: four corresponding
to $Z'\rightarrow f\bar f$ followed by
$W$ emission from one of the two (charged or neutral) fermions leaving
the $Z'$ vertex and one including the $Z'$ decay into $W^+W^-$.
In Fig. 1 we show for the $Z'\rightarrow e^- \mu^+ \rlap/p$
events and for the continuum
$WW\rightarrow e^- \mu^+ \rlap/p$
background the $e^- ,\mu^+$
(which are equal)
and $\rlap/p$
transverse momentum distributions.
For definiteness we use the $Z_{\chi}$ model with
$\Gamma_{Z'} = 0.012 \ M_{Z'}$
(assuming that the open channels are those involving only known
particles, including the top quark)
and
with a $Z'Z^0$ mixing angle
$sin \theta_3 = -\frac{0.0034}{M_{Z'}^2(TeV^2)}$ \cite{ZCHI}.
We take, for illustration, the EHLQ (set 1)
structure functions \cite{STRUC} and $\sqrt{s} = 16 \ TeV$ (LHC).
(The numerical results vary significantly for different
structure functions but do not change the conclusions.
In particular the HMRS structure functions \cite{HMRS} give a
$\sim 30 \%$ larger cross sections.)
We have
used REDUCE \cite{REDUCE} and MATHEMATICA \cite{MATHEMATICA} for calculating
the exact amplitudes and RAMBO \cite{RAMBO}
for generating the corresponding events (we work at the
parton level).
This generator is very convenient for matrix elements which
do not fluctuate too much. This is not our case, however,
for the matrix
elements we are concerned with are very much enhanced
when all the internal lines are
near on-shell. This means that we have to generate (very)
large statistics to obtain a small error.
The $Z'$ contributions from $W$ emission and from $W^+W^-$
are comparable, although model dependent.
To have a good
grip of these events
we must note that the contribution of any of the diagrams
emitting one $W$
is large
if the $W$ is on-shell and the off-shellness of the internal lepton is small.
As the fermion propagator $[M_{Z'}(M_{Z'}-2E)]^{-1}$ is large for large $E$,
where $E$ is the energy of the external lepton leaving the $Z'$
vertex in the $Z'$ rest frame,
these events have $E\sim \frac{M_{Z'}}{2}$ \cite{LANGACKER}.
Thus, the events we are interested in have at least
one fermion ($\rlap/p, e, \mu$) with a very large momentum.
The diagram with two $W$'s give a large contribution when
both gauge bosons are on-shell.

 The two neutrinos in the final state do not allow for
reconstructing the $Z'$ mass,
and we are forced to work in the transverse plane.
As a consequence and as is apparent in Fig. 1,
the strategy for isolating the $Z'$ sample
is to require large transverse momenta.
In particular we must demand a large $\rlap/p$
transverse momentum. The $e^- \rlap/p , \mu ^+ \rlap/p$
or $e^- \mu ^+$ transverse angle distributions are of no help.
Further cuts reduce
the signal without improving the signal to background ratio.
For this example
$408 \ Z'\rightarrow e \bar {\mu} \rlap/p$
events will be produced at LHC
($\int {\sl L} dt = 10^5 \ pb^{-1}$), where we sum
$e^- \mu^+ \rlap/p$ and $e^+ \mu^- \rlap/p$ events. (Universality
implies equal distributions under the interchange of $e$ and $\mu$
for a given charge assignment.) Requiring
$\rlap/p_t > 200 \ GeV, \ p_t^{e,\mu} > 50 \ GeV$,
as suggested by Fig. 1
(not very demanding pseudorapidity cuts make no difference),
$151$ events are expected for the signal.
For the same cuts the expected number of events for
the $WW$ continuum background is $23$.
We are now ready to discuss the significance of these processes.

 A new $Z'$ of the usual ($E_6$) type, for instance, $Z_{\chi}$
with a mass $\sim 1 \ TeV$ will
produce a sample of $60,000/30,000$
$e\bar e, \mu\bar{\mu}$ pairs
at LHC/SSC (we assume $\int {\sl L} dt = 10^5/10^4 \ pb^{-1}$
for LHC/SSC).
These events should allow for the measurement of $M_{Z'}$.
The question is whether four fermion $Z'$ decays are observable
at all.
$Z'$ decays involving jets, for instance
$Z'\rightarrow q\bar qW, q\bar qZ$, may not be observable due to
the large, irreducible, QCD background.
(However, if the techniques suggested in Ref. [12] can be
used to observe $Z'\rightarrow W (\rightarrow
l\bar {\nu}) W (\rightarrow jj)$, then reconstruction of $M_{Z'}$
can be done for true semileptonic $Z'$ decays and the situation
would be much better; we do not consider this analysis here.)
The $Z'\rightarrow
l\bar lZ, \nu \bar {\nu} Z\rightarrow l\bar l\nu \bar {\nu}$ decays have
small cross sections and after cuts it would be difficult to make
any definite statement ($l=e,\mu$).
This makes the $Z'\rightarrow
l\bar {\nu} W, W^+W^- \rightarrow l\nu \bar l\bar {\nu}$
decay the most promising leptonic one. However,
even this mode seems unobservable for
$M_{Z'} > 2 \ TeV$.
(Photon emission is of no interest
here for it does not make any difference among extended electroweak
models.)
In Table 1 we give the minimal width, the (largest)
$Z'Z^0$ mixing angle $sin \theta _3$
and the $Z'$ couplings to leptons for the popular models
$\chi, LR, \psi, \eta$ \cite{ZP,PDB}.
(Note that $g_{Z'\nu_{lL}\nu_{lL}} = g_{Z'l_Ll_L}$
as required by gauge invariance,
for both fermions
belong to the same ($SU(2)_L$) multiplet.)
We also quote the leptonic factors entering into the
forward-backward asymmetry,
$\frac{x^2 - 1}{x^2 + 1}, \
x^2\equiv\frac{g^2_{Z'l_Ll_L}}{g^2_{Z'l_Rl_R}}$,
and into the ratio of the $Z' \rightarrow e\bar {\mu} \rlap/p$
to the $Z' \rightarrow l \bar l$
cross sections,
$\frac{x^2}{x^2 + 1}$.
The latter is equal to the former plus $1$ and divided by $2$.
In Table 2 we give for LHC, SSC and for
$M_{Z'} = 1$ (upper values), $1.5$ (lower values) $TeV$
the
$pp \rightarrow \gamma, Z, Z' \rightarrow e\bar e, \mu \bar {\mu}$
cross section around the $Z'$ peak
(where the contribution of the standard model is at most several
per cent the $Z'$ one),
the integrated forward-backward asymmetry and
the $Z' \rightarrow e\bar {\mu}\rlap/p$ cross section
after cuts (in parentheses we quote the
$Z' \rightarrow l\bar {\nu}W$ contribution, corresponding
to $sin \theta _3 = 0$).
We demand
$\rlap/p_t > 200 \ (250) \ GeV, \ p_t^{e,\mu} > 50 \ GeV$
for $M_{Z'} = 1 (1.5) \ TeV$.
(We use the EHLQ, set 1, structure functions. The HMRS
structure functions give $\sim 20 - 30 \%$ larger
cross sections and $\sim 10 - 20 \%$ smaller asymmetries.
All numbers quoted, and in particular $\sigma _{ e\bar {\mu}\rlap/p }$,
have statistical
errors which can be as large as $30 \%$, depending on the model and
on the $M_{Z'}$ mass. We have generated typically $5$ million events
per case.)
Thus, if a new $Z'$ exists with a mass $\sim 1 \ TeV$ some
$Z' \rightarrow e\bar {\mu}\rlap/p$ events should be
detected at LHC and SSC. With the same cuts the
continuum $WW$ cross sections are
$0.23 (0.09) \ /1.06 (0.48) \ fb$ at LHC/SSC,
where the numbers in parentheses correspond
to $\rlap/p_t > 250 \ GeV$.

 As can be seen in Table 2 the
$Z'\rightarrow
l\bar {\nu} W, W^+W^- \rightarrow e\bar {\mu}\rlap/p
\ (Z'\rightarrow  l \bar {\nu}W \rightarrow e\bar {\mu}\rlap/p)$
cross sections are model dependent.
The $Z' \rightarrow e\bar {\mu}\rlap/p$ cross sections
with no cuts are, for instance for $M_{Z'} = 1 \ TeV$ and
for LHC, $\chi : \ 4.08 (3.22) \ fb; LR : \ 5.34 (1.48) \ fb;
\psi : \ 2.55 (0.90) \ fb; \eta : \ 3.47 (0.42) \ fb$
(the numbers in parentheses correspond to
$sin \theta _3 = 0$).
By comparing these cross sections to those in Table 2 we observe
that the cut on $\rlap/p$ increases the ratio of $W$ emission
to $W^+W^-$ cross sections. At any rate, the contributions
of the $W$ emission and of $W^+W^-$ are comparable and
indistinguishable after cuts. The $e\bar {\mu}\rlap/p$
sample constrains the model although it does not allow
for an independent measurement of the $W$ emission and the
$W^+W^-$ contributions. The former is proportional to the
$Z'$ coupling to left-handed leptons,
$g_{Z'\nu_{lL}\nu_{lL}} = g_{Z'l_Ll_L}$
(as required by
$SU(2)_L$ invariance);
and the latter to the $Z'Z^0$ mixing angle
$sin \theta _3$ so it will be important to
eventually separate them. The $W$ emission cross section
(in parentheses in Table 2) normalized to the
$Z' \rightarrow l\bar l$ cross section (in the
first columns in Table 2) is proportional to
$\frac{x^2}{x^2 + 1}$ in Table 1.
Whereas the $Z' \rightarrow W^+W^-$ cross section
is proportional to $sin ^2\theta _3$ in the same Table.
$sin \theta _3$ can be only measured in this process. $x^2$, however,
is also related to the forward-backward asymmetry.
This forward-backward asymmetry, which will be measured with
a higher precision, has a more involved model dependence
as can be seen comparing the $\frac{x^2 - 1}{x^2 + 1}$
column in Table 1 with the $A_{FB}$ columns in Table 2.
$A_{FB}$ is also proportional to a similar factor involving the
$Z'$ couplings to quarks. Besides, up and down quark
contributions must be summed up and the rapidity dependence
integrated.
At any rate, if a new $Z'$ exists with a mass $\sim 1 \ TeV$, the
$Z' \rightarrow e\bar {\mu}\rlap/p$ decays
can help to distinguish among different models in a
way complementary to the information coming from the forward-backward
asymmetry.

\vskip 1cm

\noindent
{\bf Acknowledgements}
\vskip 0.5cm

 We thank F. Cornet, G. Kane, P. Langacker and J. Vidal for discussions.

\vskip 2.5cm

\noindent
{\bf Table Captions}
\vskip 0.5cm

\noindent
{\it Table 1.}
Minimal width, (largest)
$Z'Z^0$ mixing angle, and leptonic $Z'$ couplings for
$Z_{\chi, LR, \psi, \eta}$. The leptonic factors entering into the
forward-backward asymmetry and into the ratio
$\frac{\sigma (Z' \rightarrow e\bar{\mu}\rlap/p)}
{\sigma (Z'\rightarrow l\bar l)}$
are also given.
$(x^2\equiv\frac{g^2_{Z'l_Ll_L}}{g^2_{Z'l_Rl_R}}.)$
\vskip 0.25cm

\noindent
{\it Table 2.}
Total cross section for $Z'$ decay into lepton ($e \bar{e},
\mu \bar{\mu}$) pairs, integrated forward-backward asymmetry and the
$Z' \rightarrow e\bar{\mu}\rlap/p$ cross section
after cuts, $\rlap/p_t > 200(250) \ GeV,\ p^{e,\mu}_t > 50 \ GeV$
for $M_{Z'} = 1 (1.5) \ TeV$,
for LHC and SSC. The upper (lower) values correspond to $M_{Z'}
= 1 (1.5) \ TeV$.
The numbers in parentheses correspond to the
$Z' \rightarrow l\bar {\nu} W\rightarrow e\bar{\mu}\rlap/p$
contribution, corresponding to $sin \theta _3 = 0$.
\vskip 1.5cm

\noindent
{\bf Figure Captions}
\vskip 0.5cm

\noindent
{\it Fig. 1.}
Transverse momentum distributions for
$Z' \rightarrow e^-\mu^+\rlap/p$
(solid curves) and the
continuum $WW$ background (dashed curves). The
charged lepton distributions are the same for $e^-$ and $\mu^+$.

\newpage

\leftskip=-1cm
\begin{tabular}{|c|cccccc|}
\hline
\multicolumn{1}{|c}{} & $\frac{\Gamma_{Z'}}{M_{Z'}}$ &
$sin \theta_3 M^2_{Z'} (TeV^2)$ & $g_{Z'l_Ll_L}\frac{c_W}{e}$
& $g_{Z'l_Rl_R}\frac{c_W}{e}$
& $\frac{x^2-1}{x^2+1}$ &
\multicolumn{1}{c|}{$\frac{x^2}{x^2+1}$} \\
\hline
$\chi$ & $0.012$ &
$-0.0034$ & $\frac{1}{2}\sqrt{\frac{3}{2}}$ &
$\frac{1}{2}\sqrt{\frac{1}{6}}$ & $\frac{8}{10}$ &
$\frac{9}{10}$ \\
$LR$ & $0.021$ &
$-0.0063$ & $\frac{1}{2}\frac{s_W}{\sqrt{1-2s^2_W}}$ &
$\frac{1}{2}\frac{-1+3s^2_W}{s_W\sqrt{1-2s^2_W}}$ & $-0.290$ &
$0.355$ \\
$\psi$ & $0.006$ &
$-0.0043$ & $-\frac{1}{6}\sqrt{\frac{5}{2}}$ &
$\frac{1}{6}\sqrt{\frac{5}{2}}$ & $0$ &
$\frac{1}{2}$ \\
$\eta$ & $0.007$ &
$-0.0055$ & $\frac{1}{6}$ &
$\frac{1}{3}$ & $-\frac{3}{5}$ &
$\frac{1}{5}$ \\
\hline
\end{tabular}
\vskip .5cm
\centerline{Table 1}
\vskip 1.5cm

\begin{tabular}{|cccccccc|}
\hline
\multicolumn{1}{|c}{} &  &
LHC &  &  &
 & SSC & \multicolumn{1}{c|}{} \\
\multicolumn{1}{|c}{} & $\sigma_{l\bar l}(pb)$ &
$A_{FB}$ & $\sigma _{e \bar {\mu} \rlap/p}(fb)$ &  &
$\sigma_{l\bar l}(pb)$ & $A_{FB}$ &
\multicolumn{1}{c|}{$\sigma _{e \bar {\mu} \rlap/p}(fb)$} \\
\hline
$\ \chi \ \ \Bigl\{$ & $\begin{array}{c} 0.63 \\ 0.12 \end{array}$ &
$\begin{array}{c} -0.12 \\ -0.13 \end{array}$ &
$\begin{array}{c} 1.51 (1.36) \\ 0.52 (0.46) \end{array}$ &  &
$\begin{array}{c} 2.82 \\ 0.69 \end{array}$ &
$\begin{array}{c} -0.10 \\ -0.11 \end{array}$ &
$\begin{array}{c} 6.75 (6.08) \\ 3.08 (2.76) \end{array}$ \\
$LR \Bigl\{$ & $\begin{array}{c} 0.74 \\ 0.14 \end{array}$ &
$\begin{array}{c} 0.10 \\ 0.11 \end{array}$ &
$\begin{array}{c} 1.11 (0.62) \\ 0.44 (0.22) \end{array}$ &  &
$\begin{array}{c} 3.11 \\ 0.78 \end{array}$ &
$\begin{array}{c} 0.09 \\ 0.10 \end{array}$ &
$\begin{array}{c} 5.83 (2.62) \\ 2.60 (1.26) \end{array}$ \\
$\ \psi \ \ \Bigl\{$ & $\begin{array}{c} 0.29 \\ 0.06 \end{array}$ &
$\begin{array}{c} 0.01 \\ 0.01 \end{array}$ &
$\begin{array}{c} 0.65 (0.38) \\ 0.26 (0.14) \end{array}$ &  &
$\begin{array}{c} 1.20 \\ 0.30 \end{array}$ &
$\begin{array}{c} 0.01 \\ 0.01 \end{array}$ &
$\begin{array}{c} 2.66 (1.56) \\ 1.35 (0.72) \end{array}$ \\
$\ \eta \ \ \Bigl\{$ & $\begin{array}{c} 0.35 \\ 0.07 \end{array}$ &
$\begin{array}{c} -0.01 \\ -0.01 \end{array}$ &
$\begin{array}{c} 0.66 (0.18) \\ 0.28 (0.06) \end{array}$ &  &
$\begin{array}{c} 1.40 \\ 0.36 \end{array}$ &
$\begin{array}{c} -0.01 \\ -0.01 \end{array}$ &
$\begin{array}{c} 3.47 (0.74) \\ 1.09 (0.34) \end{array}$ \\
\hline
\end{tabular}
\vskip .5cm
\centerline{Table 2}

\end{document}